\documentclass[aps,prl,twocolumn,showpacs,superscriptaddress,groupedaddress]{revtex4}  
\usepackage{graphicx}    
\usepackage{dcolumn}    
\usepackage{bm}              
\usepackage{amssymb}   

\newcommand{\Zprime}	    {$Z'$}

\newcommand{\about}          {$\sim$}
\newcommand{\pythia}        {\textsc{pythia}}

\begin{document}

 \hspace{5.2in} \mbox{Fermilab-Pub-10-300-E}

\title{Search for a heavy neutral gauge boson in the dielectron channel \\
	with 5.4~fb$^{-1}$ of $\mathbf{p\bar{p}}$ collisions at $\mathbf{\sqrt{s} = 1.96}$~TeV}

\affiliation{Universidad de Buenos Aires, Buenos Aires, Argentina}
\affiliation{LAFEX, Centro Brasileiro de Pesquisas F{\'\i}sicas, Rio de Janeiro, Brazil}
\affiliation{Universidade do Estado do Rio de Janeiro, Rio de Janeiro, Brazil}
\affiliation{Universidade Federal do ABC, Santo Andr\'e, Brazil}
\affiliation{Instituto de F\'{\i}sica Te\'orica, Universidade Estadual Paulista, S\~ao Paulo, Brazil}
\affiliation{Simon Fraser University, Vancouver, British Columbia, and York University, Toronto, Ontario, Canada}
\affiliation{University of Science and Technology of China, Hefei, People's Republic of China}
\affiliation{Universidad de los Andes, Bogot\'{a}, Colombia}
\affiliation{Charles University, Faculty of Mathematics and Physics, Center for Particle Physics, Prague, Czech Republic}
\affiliation{Czech Technical University in Prague, Prague, Czech Republic}
\affiliation{Center for Particle Physics, Institute of Physics, Academy of Sciences of the Czech Republic, Prague, Czech Republic}
\affiliation{Universidad San Francisco de Quito, Quito, Ecuador}
\affiliation{LPC, Universit\'e Blaise Pascal, CNRS/IN2P3, Clermont, France}
\affiliation{LPSC, Universit\'e Joseph Fourier Grenoble 1, CNRS/IN2P3, Institut National Polytechnique de Grenoble, Grenoble, France}
\affiliation{CPPM, Aix-Marseille Universit\'e, CNRS/IN2P3, Marseille, France}
\affiliation{LAL, Universit\'e Paris-Sud, CNRS/IN2P3, Orsay, France}
\affiliation{LPNHE, Universit\'es Paris VI and VII, CNRS/IN2P3, Paris, France}
\affiliation{CEA, Irfu, SPP, Saclay, France}
\affiliation{IPHC, Universit\'e de Strasbourg, CNRS/IN2P3, Strasbourg, France}
\affiliation{IPNL, Universit\'e Lyon 1, CNRS/IN2P3, Villeurbanne, France and Universit\'e de Lyon, Lyon, France}
\affiliation{III. Physikalisches Institut A, RWTH Aachen University, Aachen, Germany}
\affiliation{Physikalisches Institut, Universit{\"a}t Freiburg, Freiburg, Germany}
\affiliation{II. Physikalisches Institut, Georg-August-Universit{\"a}t G\"ottingen, G\"ottingen, Germany}
\affiliation{Institut f{\"u}r Physik, Universit{\"a}t Mainz, Mainz, Germany}
\affiliation{Ludwig-Maximilians-Universit{\"a}t M{\"u}nchen, M{\"u}nchen, Germany}
\affiliation{Fachbereich Physik, Bergische  Universit{\"a}t Wuppertal, Wuppertal, Germany}
\affiliation{Panjab University, Chandigarh, India}
\affiliation{Delhi University, Delhi, India}
\affiliation{Tata Institute of Fundamental Research, Mumbai, India}
\affiliation{University College Dublin, Dublin, Ireland}
\affiliation{Korea Detector Laboratory, Korea University, Seoul, Korea}
\affiliation{CINVESTAV, Mexico City, Mexico}
\affiliation{FOM-Institute NIKHEF and University of Amsterdam/NIKHEF, Amsterdam, The Netherlands}
\affiliation{Radboud University Nijmegen/NIKHEF, Nijmegen, The Netherlands}
\affiliation{Joint Institute for Nuclear Research, Dubna, Russia}
\affiliation{Institute for Theoretical and Experimental Physics, Moscow, Russia}
\affiliation{Moscow State University, Moscow, Russia}
\affiliation{Institute for High Energy Physics, Protvino, Russia}
\affiliation{Petersburg Nuclear Physics Institute, St. Petersburg, Russia}
\affiliation{Stockholm University, Stockholm and Uppsala University, Uppsala, Sweden }
\affiliation{Lancaster University, Lancaster LA1 4YB, United Kingdom}
\affiliation{Imperial College London, London SW7 2AZ, United Kingdom}
\affiliation{The University of Manchester, Manchester M13 9PL, United Kingdom}
\affiliation{University of Arizona, Tucson, Arizona 85721, USA}
\affiliation{University of California Riverside, Riverside, California 92521, USA}
\affiliation{Florida State University, Tallahassee, Florida 32306, USA}
\affiliation{Fermi National Accelerator Laboratory, Batavia, Illinois 60510, USA}
\affiliation{University of Illinois at Chicago, Chicago, Illinois 60607, USA}
\affiliation{Northern Illinois University, DeKalb, Illinois 60115, USA}
\affiliation{Northwestern University, Evanston, Illinois 60208, USA}
\affiliation{Indiana University, Bloomington, Indiana 47405, USA}
\affiliation{Purdue University Calumet, Hammond, Indiana 46323, USA}
\affiliation{University of Notre Dame, Notre Dame, Indiana 46556, USA}
\affiliation{Iowa State University, Ames, Iowa 50011, USA}
\affiliation{University of Kansas, Lawrence, Kansas 66045, USA}
\affiliation{Kansas State University, Manhattan, Kansas 66506, USA}
\affiliation{Louisiana Tech University, Ruston, Louisiana 71272, USA}
\affiliation{University of Maryland, College Park, Maryland 20742, USA}
\affiliation{Boston University, Boston, Massachusetts 02215, USA}
\affiliation{Northeastern University, Boston, Massachusetts 02115, USA}
\affiliation{University of Michigan, Ann Arbor, Michigan 48109, USA}
\affiliation{Michigan State University, East Lansing, Michigan 48824, USA}
\affiliation{University of Mississippi, University, Mississippi 38677, USA}
\affiliation{University of Nebraska, Lincoln, Nebraska 68588, USA}
\affiliation{Rutgers University, Piscataway, New Jersey 08855, USA}
\affiliation{Princeton University, Princeton, New Jersey 08544, USA}
\affiliation{State University of New York, Buffalo, New York 14260, USA}
\affiliation{Columbia University, New York, New York 10027, USA}
\affiliation{University of Rochester, Rochester, New York 14627, USA}
\affiliation{State University of New York, Stony Brook, New York 11794, USA}
\affiliation{Brookhaven National Laboratory, Upton, New York 11973, USA}
\affiliation{Langston University, Langston, Oklahoma 73050, USA}
\affiliation{University of Oklahoma, Norman, Oklahoma 73019, USA}
\affiliation{Oklahoma State University, Stillwater, Oklahoma 74078, USA}
\affiliation{Brown University, Providence, Rhode Island 02912, USA}
\affiliation{University of Texas, Arlington, Texas 76019, USA}
\affiliation{Southern Methodist University, Dallas, Texas 75275, USA}
\affiliation{Rice University, Houston, Texas 77005, USA}
\affiliation{University of Virginia, Charlottesville, Virginia 22901, USA}
\affiliation{University of Washington, Seattle, Washington 98195, USA}
\author{V.M.~Abazov} \affiliation{Joint Institute for Nuclear Research, Dubna, Russia}
\author{B.~Abbott} \affiliation{University of Oklahoma, Norman, Oklahoma 73019, USA}
\author{M.~Abolins} \affiliation{Michigan State University, East Lansing, Michigan 48824, USA}
\author{B.S.~Acharya} \affiliation{Tata Institute of Fundamental Research, Mumbai, India}
\author{M.~Adams} \affiliation{University of Illinois at Chicago, Chicago, Illinois 60607, USA}
\author{T.~Adams} \affiliation{Florida State University, Tallahassee, Florida 32306, USA}
\author{G.D.~Alexeev} \affiliation{Joint Institute for Nuclear Research, Dubna, Russia}
\author{G.~Alkhazov} \affiliation{Petersburg Nuclear Physics Institute, St. Petersburg, Russia}
\author{A.~Alton$^{a}$} \affiliation{University of Michigan, Ann Arbor, Michigan 48109, USA}
\author{G.~Alverson} \affiliation{Northeastern University, Boston, Massachusetts 02115, USA}
% ...
%
%
\collaboration{The D0 Collaboration\footnote{with visitors from
$^{a}$Augustana College, Sioux Falls, SD, USA,
$^{b}$The University of Liverpool, Liverpool, UK,
$^{c}$SLAC, Menlo Park, CA, USA,
$^{d}$ICREA/IFAE, Barcelona, Spain,
$^{e}$Centro de Investigacion en Computacion - IPN, Mexico City, Mexico,
$^{f}$ECFM, Universidad Autonoma de Sinaloa, Culiac\'an, Mexico,
and 
$^{g}$Universit{\"a}t Bern, Bern, Switzerland.%
}} \noaffiliation
\vskip 0.25cm
       
\date{August 11, 2010}

\begin{abstract}
We report the results of a search for a heavy neutral gauge boson \Zprime\ decaying into the dielectron final state using data corresponding to an integrated luminosity of 5.4~fb$^{-1}$ collected by the D0 experiment at the Fermilab Tevatron Collider.  No significant excess above the standard model prediction is observed in the dielectron invariant-mass spectrum.  We set 95\% C.L. upper limits on
$\sigma \left ( p\bar{p} \rightarrow Z' \right ) \times BR(Z' \rightarrow ee)$ depending on the dielectron invariant mass.  These cross section limits are used to determine lower mass limits for \Zprime\ bosons in a variety of models. For the sequential standard model \Zprime\ boson a lower mass limit of 1023~GeV is obtained.
\end{abstract}

\maketitle


The gauge group structure of the standard model (SM), $SU(3)_C \otimes SU(2)_L \otimes U(1)_Y$, can be extended with an additional $U(1)$ group, which may arise in models derived from grand unified theories (GUT) that are based on groups with rank larger than four~\cite{Leike:1998wr}. Additional $U(1)$ groups can also arise from higher dimensional constructions like string compactifications. In many models of GUT symmetry breaking, $U(1)$ groups survive at relatively low energies, leading to corresponding neutral gauge bosons, commonly referred to as \Zprime\ bosons~\cite{langacker}. Such \Zprime\ bosons typically couple to SM fermions via the electroweak interaction, and can be observed at hadron colliders as narrow resonances through the process $q \bar{q} \rightarrow Z' \rightarrow e^+e^-$.
There is no simple general parametrization that can be applied to all the \Zprime\ models.  Nevertheless, the models can be distinguished according to the strength of the gauge coupling, $g_{Z'}$, for the additional $U(1)$ group. The models with coupling of electroweak strength are called canonical. The sequential standard model (SSM) \Zprime\ boson is a canonical example, where the SSM \Zprime\ boson ($Z'_{SSM}$) is defined to have the same couplings as the SM $Z$ boson. The SSM \Zprime\ boson is often used as benchmark~\cite{langacker, cvetic}. An additional example of a canonical model can be derived from the superstring inspired $E_6$ models~\cite{e6models}.  
The decomposition of $E_6$ can give rise to two additional $U(1)$ factors through $E_6 \rightarrow SO(10) \times U(1)_\psi$ and $SO(10) \rightarrow SU(5) \times U(1)_\chi$. These groups are associated with the gauge fields $Z'_\psi$ and $Z'_\chi$ that can mix and, at the TeV scale, can give rise to additional \Zprime\ bosons through the linear combination
\begin{equation}
	Z'(\theta) = Z'_\chi \sin \theta + Z'_\psi \cos \theta,
\end{equation}
where $0 \leq \theta < \pi$ is a mixing angle~\cite{rosner}. 
The most commonly referenced \Zprime\ boson models arising from $E_6$ are summarized in Table~\ref{tab:e6_models}~\cite{pythia_couplings}.

\begin{table}
\caption{A selection of commonly used $E_6$ models~\cite{pythia_couplings}.}
\begin{ruledtabular}
\begin{tabular}{ccc}
  Model     & $\sin \theta$   & $\cos \theta$ \\  \hline
 $Z'_\psi$ &      0                  &       1               \\
 $Z'_\chi$ &      1                  &       0               \\ 
 $Z'_\eta$ & $\sqrt{3/8}$    & $\sqrt{5/8}$   \\
 $Z'_I$       & $\sqrt{5/8}$    & $-\sqrt{3/8}$   \\
 $Z'_{sq}$ & $3\sqrt{6}/8$  & $-\sqrt{10}$/8   \\
 $Z'_N$     & $1/4$              & $-\sqrt{15}/4$   \\

\end{tabular}
\end{ruledtabular}
\label{tab:e6_models}
\end{table}

An example of a non-canonical model is the $U(1)_X$ Stueckelberg extension of the standard model (StSM) that gives rise to a very narrow \Zprime\ boson~\cite{kors, nath}. The Stueckelberg mechanism allows for the possibility of an Abelian gauge boson to gain mass without the requirement of a Higgs mechanism.  The new parameters that are introduced in this model are the StSM mass mixing parameter, $\epsilon$, and the \Zprime\ boson mass, $M_{Z'}$. In the limit $\epsilon \rightarrow 0$, the Stueckelberg sector decouples from the SM~\cite{feldman}.

In this Letter, we report on a search for a \Zprime\ boson decaying into an electron pair with the D0 detector at the Fermilab Tevatron Collider, where protons and antiprotons collide at $\sqrt{s} = 1.96$~TeV. A \Zprime\ boson would appear as a narrow resonance in the $ee$ invariant mass spectrum, with a natural width smaller than the resolution of the D0 electromagnetic calorimeter. A previous Tevatron search by the CDF collaboration~\cite{cdf_25}, corresponding to 2.5~fb$^{-1}$ of integrated luminosity, sets a lower mass limit on SSM \Zprime\ bosons of 963~GeV and reports a discrepancy over the expected SM background at $M_{ee} \sim 240$~GeV equivalent to 2.5 standard deviations. 
The CDF collaboration has also performed a search in the $Z' \rightarrow \mu\mu$ channel~\cite{cdf_dimuon}, corresponding to 2.3~fb$^{-1}$ of integrated luminosity, with 95\% C.L. upper limits on $\sigma \left ( p\bar{p} \rightarrow Z' \right ) \times BR(Z' \rightarrow \mu \mu)$ ranging from $\sim$50~fb to $\sim$3.2~fb for $M_{Z'}$ between 175~GeV and 1100~GeV.


The D0 detector~\cite{d0det} is composed of a central tracking system surrounded by a 2~T superconducting solenoidal magnet and a central preshower detector (CPS), a calorimeter, and a muon spectrometer.  The central tracking system includes a silicon microstrip tracker (SMT) and a scintillating fiber tracker (CFT) that are designed to provide coverage for particles in the pseudorapidity range $|\eta| < 3$, where $\eta = -\mathrm{ln}\left [ \mathrm{tan} \left ( \theta / 2 \right ) \right ]$, and $\theta$ is the polar angle with respect to the proton beam direction.  The azimuthal angle is denoted by $\phi$.  The CPS is located between the solenoid and the inner layer of the central calorimeter and is formed of approximately one radiation length of lead absorber followed by three layers of scintillating strips. The calorimeter consists of a central section (CC) covering $|\eta| \lesssim 1.1$ and two end calorimeters (EC) that extend the EM coverage to $\eta \approx  4.1$, with all three sections housed in separate cryostats~\cite{d0cal}.  Each section consists of an inner electromagnetic (EM) section, and an outer hadronic.  The EM calorimeter is segmented into four longitudinal layers (EM$i$, $i = 1,..., 4$) with transverse segmentation of $\Delta \eta \times \Delta \phi = 0.1 \times 0.1$, except for the finely segmented third layer where it is $0.05 \times 0.05$. The muon system, covering $|\eta| < 2$, is located beyond the calorimeter and is composed of a layer of tracking detectors and scintillation trigger counters in front of 1.8~T iron toroidal magnets, and followed by two similar layers after the toroids.  The luminosity is measured using plastic scintillator arrays in front of the end calorimeters. The data acquisition system includes a three-level trigger, designed to accommodate the high instantaneous luminosity. The data sample was collected between July 2002 and June 2009 using triggers requiring at least two clusters of energy deposits in the EM calorimeter and corresponds to an integrated luminosity of 5.4 $\pm$ 0.3~fb$^{-1}$~\cite{d0lumi}. 

The event selection requires two isolated electron candidates in the central section of the calorimeter. An electron candidate is characterized by an EM cluster with transverse momentum $p_T > 25$~GeV and $|\eta| < 1.1$, reconstructed in a cone of radius ${\cal R} = \sqrt{ \left ( \Delta \eta \right )^2 + \left ( \Delta \phi \right )^2} = 0.4$. At least 97\% of the EM cluster energy must be deposited in the EM section of the calorimeter and its energy must be isolated in the calorimeter,  $\left [E_\mathrm{tot} \left ( 0.4 \right ) - E_\mathrm{EM} \left ( 0.2 \right ) \right ] /  E_\mathrm{EM} \left ( 0.2 \right ) < 0.07$, where $E_\mathrm{tot} \left ( { \cal R} \right )$ and $E_\mathrm{EM} \left ( {\cal R} \right )$ are the total energy and the energy in the EM section, respectively, within a cone of radius ${\cal R}$ around the electron direction. In addition, the EM cluster is required to be consistent with an electron shower shape, using a $\chi^2$ test and a neural network discriminant~\cite{d0photonres}.
The EM cluster is required to be spatially matched to either a reconstructed track or a pattern of hits in the SMT and CFT consistent with the passage of an electron. The scalar sum of the $p_T$ of all tracks originating from the $p \bar{p}$ interaction vertex (PV) in an annulus of $0.05 < {\cal R} < 0.4$ around the cluster is required to be less than 2.5~GeV.
Events are only considered if the PV lies within 60~cm of the geometrical center of the detector in the coordinate along the beam axis to be fully within the SMT acceptance. 
The two electron candidates are not required to have opposite charges to avoid losses due to charge misidentification. The data sample consists of 185,264 events that satisfy these selection criteria in the dielectron invariant mass control region $60 < M_{ee} < 150$~GeV and 1332 events in the search region $M_{ee} > 150$~GeV.

Signal and SM background events are generated using \pythia~\cite{pythia} with the CTEQ6L1~\cite{cteq} parametrization of the parton distribution functions (PDFs), and processed through the D0 detector simulation based on {\sc geant3}~\cite{geant} adding zero bias events, and the same reconstruction software as the data. Signal templates based on the SSM \Zprime\ boson have been generated up to masses of 1100~GeV. The width of the resonance scales with the \Zprime\ boson mass, according to $\Gamma_{Z'} = \Gamma_Z \times M_{Z'}/M_Z$, where $M_Z$  and $\Gamma_Z$ are the mass and width of the $Z$ boson. For $M_{Z'} \geq 2 m_t$ the decay channel to top quarks opens up, thus increasing the width of the resonance. The signal selection efficiency increases from \about 22\% to \about 44\% for $M_{Z'}$ between 175 and 1100~GeV independent of the type of \Zprime\ boson discussed in this Letter.

The dominant irreducible background is due to the Drell-Yan (DY) process. A mass-dependent k-factor~\cite{kfactor} has been applied to the \pythia\ dielectron invariant mass spectrum to account for next-to-next-to-leading order (NNLO) contributions. The main instrumental background originates from the misidentification of one or two jets as electrons. The shape of the invariant mass spectrum for this background is obtained from data by selecting events where the EM clusters fail the $\chi^2$ test. Other SM backgrounds include $Z/\gamma^* \rightarrow \tau\tau$, $W+\gamma$, $WW$, $ZZ$, $WZ$, $W+$~jets, $t \bar{t}$, and $\gamma\gamma$ production. The contribution of these background processes is small (\about 0.6\%) and is estimated using \pythia\ corrected for higher order contributions~\cite{campbell, baur, kidonakis}.

The normalization of the various background contributions is determined by fitting the invariant mass spectrum of the data to a superposition of the backgrounds in a control region around the $Z$ boson mass (60~$< M_{ee} <$~150~GeV), where the existence of \Zprime\ bosons has been excluded by previous searches~\cite{lep}. The total number of background events in that region is fixed to the number of events that have been observed in the data. The relative contribution from the DY process and instrumental background is a free parameter, while the contribution from the other SM processes is normalized to their theoretical cross sections. The uncertainty of the background normalization is estimated by varying both the criteria to select the instrumental background sample and the fitting range, and is 2\%.

\begin{figure}
	\centering
	\includegraphics[scale=0.53]{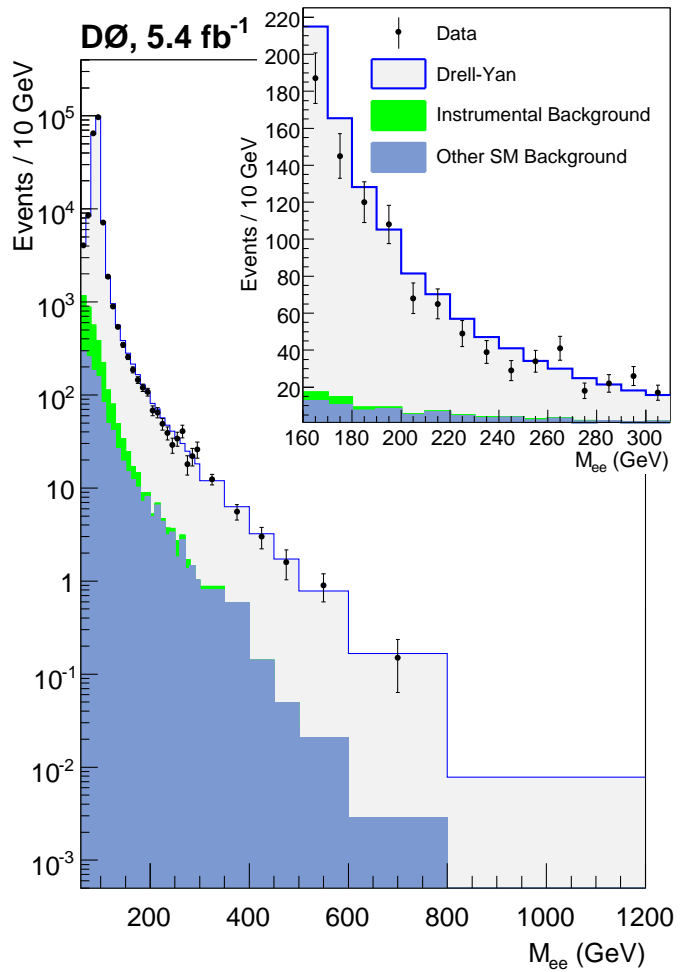}		
	\caption {Distribution of $M_{ee}$ for data, along with the total expected background for the full invariant mass range studied. Insert focuses on the area of the $M_{ee}$ spectrum from 160~GeV to 300~GeV, where the majority of observed data in the signal region lie.}
	\label{fig:inv_mass_full}
\end{figure}

Having normalized the various background contributions to data in the control region, the background shapes are used to extrapolate to higher invariant masses. The measured $ee$ invariant mass spectrum, superimposed on the expected backgrounds for the full mass range studied, is shown in Fig.~\ref{fig:inv_mass_full}. The data and expected background are generally in good agreement for the full invariant mass range studied, with a $\chi^2$ over degrees of freedom equal to $118.5 / 113$.

In the absence of a heavy resonance signal, the $ee$ invariant mass distribution is used to calculate an upper limit on the production cross section of \Zprime\ bosons multiplied by the branching ratio into the $ee$ final state, using a Poisson log-likelihood ratio (LLR) test statistics~\cite{collie}. The expected limits are calculated using the median of the LLR distribution for a background-only hypothesis. The observed limit, obtained including all the fluctuations present in the data, is expected to be contained in the $\pm1$ and $\pm2$ standard deviations region with a probability of 68\% and 95\%, respectively. An observed limit significantly outside the expected range would indicate either a poor modeling of the background or that the data is inconsistent with the background-only hypothesis. 

The following systematic uncertainties on the expected background and the signal have been considered for the limit calculation.  The uncertainties affecting the expected background include the electron identification efficiency (3.0\% per electron), the mass dependence of the DY associated NNLO k-factor (5.0\%), and the background normalization (2.0\%).
Uncertainties that affect the signal include the integrated luminosity (6.1\%), the PDFs for signal acceptance (0.4\% -- 7.6\%), 
the electron identification efficiency (3.0\% per electron), the EM cluster energy resolution (6.0\%), and the trigger efficiency (0.1\%). 
For the EM energy resolution and the background normalization, both the effects on the normalization and on the shape of the invariant mass distribution have been considered in extracting limits. For the remaining systematic sources only the changes to the overall background normalization or signal detection efficiency have been considered. The systematic uncertainties are included via convolution of the Poisson probability distributions for signal and background with Gaussian distributions corresponding to the different sources of systematic uncertainties taking into account all relevant correlations between systematics' sources.

\begin{figure}
	\centering
	\includegraphics[scale=0.46]{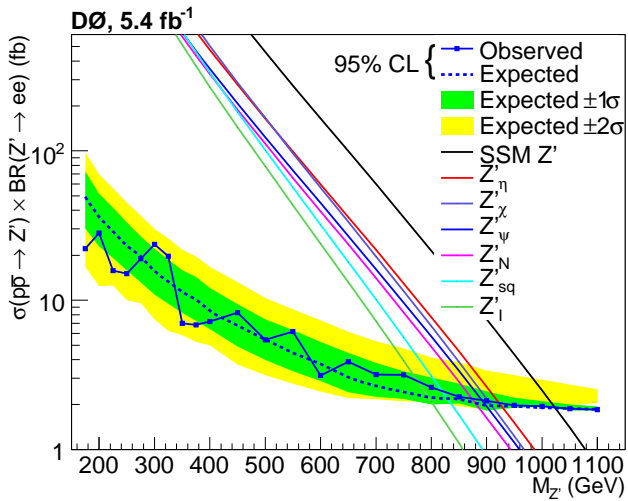}
	\caption {The observed and expected 95\% C.L. upper limits on $\sigma \left ( p\bar{p} \rightarrow Z' \right ) \times BR  \left ( Z' \rightarrow ee\right )$ as a function of $M_{Z'}$, compared to the theoretical predictions of the cross section for the SSM \Zprime\ boson and the \Zprime\ bosons arising from the $E_6$ model. The median expected limits are shown together with the $\pm1$ and $\pm2$ standard deviation bands.}
	\label{fig:limit_e6}
\end{figure}

The observed upper limits on the production cross section multiplied by the branching ratio into an $ee$ pair for the process $p \bar{p} \rightarrow Z' \rightarrow ee$ are given in Table~\ref{tab:production_limits} as a function of the mass hypothesis, together with the median expected limits calculated under the assumption that the observed dielectron invariant mass spectrum arises only from the backgrounds considered in the analysis. Figure~\ref{fig:limit_e6} shows these limits together with  the $\pm1$ and $\pm2$ standard deviation bands on the expected limit, and the cross section predictions for SSM and E$_6$ \Zprime\ bosons~\cite{pythia_couplings} where a constant k-factor of 1.3~\cite{carena} has been applied to the \pythia\ cross section. 
Since this analysis searches for a resonance instead of an enhancement in the total cross section, signal cross section predictions are calculated by integrating over the region $\left[M_{Z'} - 10 \times \Gamma_{Z'}, \infty \right]$, where $\Gamma_{Z'}$ is the width of the SSM \Zprime\ boson, thus excluding \Zprime\ boson events which do not contribute to the resonant region. For $M_{Z'} < 500$~GeV the difference between the cross section in the region defined above and the total cross section is less than 5\%, while for a $M_{Z'} = 1$~TeV SSM \Zprime\ boson it is \about 40\%.
The mass limits on the specific models of \Zprime\ bosons considered are given in Table~\ref{tab:mass_limits}.

\begin{figure}
	\centering
	\includegraphics[scale=0.46]{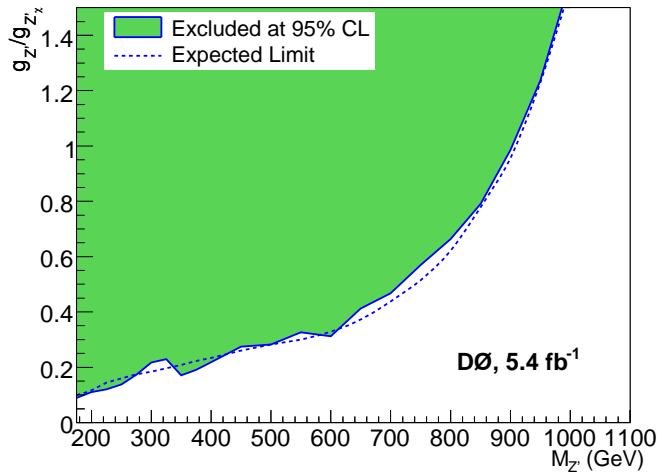}
	\caption {Excluded region in the ($M_{Z'}$, $g_{Z'}$) plane at 95\% C.L. for the $Z'_\chi$ model. The expected limit is superimposed.}
	\label{fig:couplings}
\end{figure}

These limits can be translated into upper limits on the $U(1)_{Z'}$ gauge coupling, $g_{Z'}$~\cite{pythia_couplings}, as a function of $M_{Z'}$. Figure~\ref{fig:couplings} illustrates the observed upper limits on $g_{Z'}/g_{Z'_\chi}$\footnote{Here $g_{Z'_\chi} =\sqrt{\frac{3}{8}} \cdot g \cdot  \tan{\theta_W}$, where $g = 0.626$ and $\sin{\theta_W} = \sqrt{0.23116}$.} for the $Z'_\chi$ model.

\begin{table}
\caption{Expected and observed 95\% C.L. upper limits on the production cross section multiplied by the branching ratio, $\sigma \left ( p\bar{p} \rightarrow Z' \right ) \times BR  \left ( Z' \rightarrow ee\right )$.}
\begin{ruledtabular}
\begin{tabular}{ccc}
  \Zprime\ Boson Mass & \multicolumn{2}{c}{$\sigma \left ( p\bar{p} \rightarrow Z' \right ) \times BR  \left ( Z' \rightarrow ee\right )$ (fb)}  \\
  (GeV)& 					Expected    		   & Observed				                \\ \hline
 175 & 49 & 22 \\
 200 & 36 & 28 \\
 225 & 29 & 16 \\
 250 & 23 & 15 \\
 275 & 20 & 19 \\
 300 & 16 & 24 \\
 325 & 13 & 20 \\
 350 & 11 &  7.0 \\
 375 & 10 &  6.9 \\
 400 &  8.5 &  7.2 \\
 450 &  6.8 &  8.2 \\
 500 &  5.5 &  5.4 \\
 550 &  4.4 &  6.2 \\
 600 &  3.7 &  3.1 \\
 650 &  3.1 &  3.9 \\
 700 &  2.7 &  3.2 \\
 750 &  2.4 &  3.2 \\
 800 &  2.2 &  2.6 \\
 850 &  2.2 &  2.3 \\
 900 &  2.0 &  2.1 \\
 950 &  1.9 &  2.0 \\
1000 &  1.9 &  2.0 \\
1050 &  1.9 &  1.9 \\
1100 &  1.9 &  1.9 \\
\end{tabular}
\end{ruledtabular}
\label{tab:production_limits}
\end{table}

\begin{figure}
	\centering
	\includegraphics[scale=0.46]{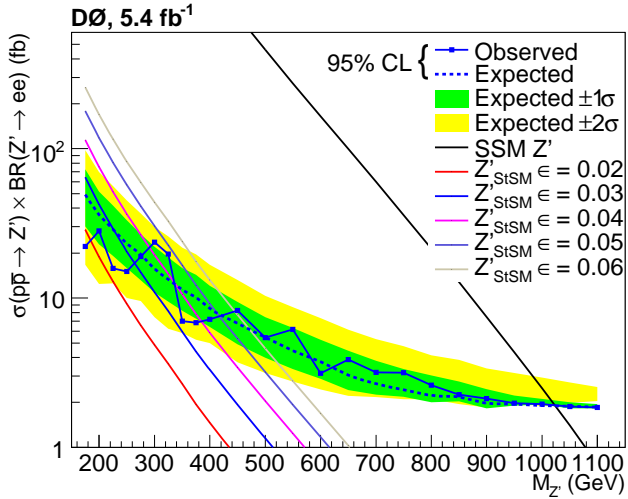}
	\caption {The observed and expected 95\% C.L. upper limits on $\sigma \left ( p\bar{p} \rightarrow Z' \right ) \times BR  \left ( Z' \rightarrow ee\right )$ as a function of $M_{Z'}$, compared to the theoretical predictions for the \Zprime\ boson cross sections in the SSM and in the StSM extension for values of $\epsilon$ ranging from 0.02 to 0.06. The median expected limits are shown together with the $\pm1$ and $\pm2$ standard deviation bands.}
	\label{fig:limit_stsm}
\end{figure}

Cross sections are calculated as a function of \Zprime\ boson mass to interpret the observed upper limits on $\sigma \left ( p\bar{p} \rightarrow Z' \right ) \times BR  \left ( Z' \rightarrow ee\right )$ as mass limits for a StSM \Zprime\ boson. 
Figure~\ref{fig:limit_stsm} shows the observed and expected limits and the cross section predictions for the StSM \Zprime\ boson for several $\epsilon$ values from 0.02 to 0.06~\cite{feldman}. The mass limits are summarized in Table~\ref{tab:mass_limits}.

\begin{table}[hbpt]
\caption{Expected and observed lower mass limits for various \Zprime\ bosons.}
\begin{ruledtabular}
\begin{tabular}{ccc}
  Model   & \multicolumn{2}{c}{Lower Mass Limit (GeV)} \\
                & Expected          & Observed               \\ \hline
$Z'_{\mathrm{SSM}}$          & 1024  & 1023 \\ 
$Z'_\eta$              &  927  &  923 \\ 
$Z'_{\chi}$            &  910  &  903 \\ 
$Z'_\psi$              &  898  &  891 \\ 
$Z'_N$                 &  879  &  874 \\ 
$Z'_{sq}$             &  829  &  822 \\ 
$Z'_I$                  &  795  &  772 \\  \hline
$Z'_{\mathrm{StSM}} (\epsilon = 0.06)$          &  471   &  443 \\ 
$Z'_{\mathrm{StSM}} (\epsilon = 0.05)$          &  414   &  417 \\ 
$Z'_{\mathrm{StSM}} (\epsilon = 0.04)$          &  340   &  289 \\ 
$Z'_{\mathrm{StSM}} (\epsilon = 0.03)$          &  227   &  264 \\ 
$Z'_{\mathrm{StSM}} (\epsilon = 0.02)$          &  ---      &  180 \\  
\end{tabular}
\end{ruledtabular}
\label{tab:mass_limits}
\end{table}

In summary, we have searched for a heavy narrow resonance in the $ee$ invariant mass spectra, using 5.4~fb$^{-1}$ of integrated luminosity collected with the D0 detector at the Fermilab Tevatron Collider. The observed spectrum agrees with the total background expected from SM processes and instrumental backgrounds. No evidence for physics beyond the SM is observed. 
For a \Zprime\ boson with SM couplings and with intrinsic width significantly smaller than the detector resolution, we set 95\% C.L. upper limits on $\sigma \left ( p\bar{p} \rightarrow Z' \right ) \times BR  \left ( Z' \rightarrow ee\right )$ between 22~fb and 1.9~fb for $M_{Z'}$ between 175~GeV and 1100~GeV. These represent the most stringent constraints to date, and translate into a lower limit on the mass of the SSM \Zprime\ boson of 1023~GeV. 

%
We thank the staffs at Fermilab and collaborating institutions,
and acknowledge support from the
DOE and NSF (USA);
CEA and CNRS/IN2P3 (France);
FASI, Rosatom and RFBR (Russia);
CNPq, FAPERJ, FAPESP and FUNDUNESP (Brazil);
DAE and DST (India);
Colciencias (Colombia);
CONACyT (Mexico);
KRF and KOSEF (Korea);
CONICET and UBACyT (Argentina);
FOM (The Netherlands);
STFC and the Royal Society (United Kingdom);
MSMT and GACR (Czech Republic);
CRC Program and NSERC (Canada);
BMBF and DFG (Germany);
SFI (Ireland);
The Swedish Research Council (Sweden);
and
CAS and CNSF (China).
%

\end{document}